# AnyDijkstra, an algorithm to compute shortest paths on images with anytime properties

Diego Ulisse Pizzagalli [1,*], Rolf Krause[2,*]

**Abstract**

Images conveniently capture the result of physical processes, representing rich source of information for data driven medicine, engineering, and science. The modeling of an image as a graph allows the application of graph-based algorithms for content analysis. Amongst these, one of the most used is the Dijkstra Single Source Shortest Path algorithm (DSSSP), which computes the path with minimal cost from one starting node to all the other nodes of the graph. However, the results of DSSSP remains unknown for nodes until they are explored. Moreover, DSSSP execution is associated to frequent jumps between distant locations in the graph, which results in non-optimal memory access, reduced parallelization, and finally increased execution time. Therefore, we propose AnyDijkstra, an iterative implementation of the Dijkstra SSSP algorithm optimized for images, that retains anytime properties while accessing memory following a cache-friendly scheme and maximizing parallelization.

[1]*Faculty of Biomedical Science, USI, Switzerland*
[2]*Euler institute, USI, Switzerland*
**\*Correspondence**: pizzad@usi.ch and rolf.krause@usi.ch

## Introduction

Finding the shortest path from one node (source) to multiple other nodes on an image is a relevant task in computer vision. Focusing on biomedical applications, the usage of shortest paths from density peaks improved the segmentation of non-convex cells in fluorescence microscopy images [1]. On light-microscopy images, Uhlman *et al.* developed a plugin for ImageJ (DiversePathsJ) to identify the skeleton of worms, following the shortest path from the head to the tail [2]. Instead, Ghidoni and colleagues reported a shortest path-based descriptor to analyze and classify medical images [3].

The Dijkstra Single Source Shortest Path (DSSSP) algorithm is one of the most used methods to computed shortest paths on a generic graph with non-negative edge costs [1]. However, images exhibit a regular structure that can be considered to optimize computation.

Let $G(V,E,w)$ be a generic, fully connected, weighted graph with $|V|$ nodes and $|E|$ edges. The naive implementation of the Dijkstra algorithm has a computational complexity of $O(|V|^2)$. This is due to the task of finding the minimum in an unsorted array of size $|V|$, which has a complexity $O(|V|)$ and is repeated for all the nodes of the graph. By using queues such as min-heap or Fibonacci heaps, the amortized computational complexity of the algorithm on a generic graph can be reduced to $O(|E|+|V|*\log(|V|))$.

However, this theoretical computational complexity does not count for data locality which has a huge impact on the time required for accessing memory.

Indeed, the Dijkstra SSSP algorithm successively selects a node to explore, as the node that is reached with the minimum cost and is not explored yet.

This rule can introduce jumping from one node to another arbitrary node. Considering the storage of the graph as an array on memory, this can introduce cache-miss. Moreover, parallel implementations of the algorithm that assign partitions of the graph to different cores might require communication to handle these jumps or more advanced strategies as reviewed in [4].

Heuristics such as A* [5] and the Jump Point Search [6] allowed to speed-up the computation by one order of magnitude on generic graphs and two orders of magnitude on regular grids, by avoiding to explore unnecessary nodes.

Similarly, we propose an efficient implementation of the Dijkstra SSSP algorithm for regular lattices, based on the intuition that a target node $t$ can be reached from a source node $s$ only by crossing horizontal and vertical edges. Therefore, we evaluated the possibility of computing shortest paths, by evaluating independently orthogonal directions and by iteratively updating the results.

The proposed strategy brings two advantages. Firstly, from a memory access point of view, it accesses nodes sequentially, which is more cache-friendly strategy than accessing elements randomly. This is relevant when implementing the algorithm as the edge cost of rectangular lattices with HxW nodes can be stored on memory on two 2D matrices of size (H-1, W) for vertical edges, and size (H, W-1) for horizontal edges. Moreover, this confers to the algorithm an anytime property that allows its interruption after n ¿= 2 iterations to obtain an approximated computation of the shortest paths. Secondly, at each step, different columns can be assigned to different processors without any need for communication. Once all the columns have been updated, results will be updated along the rows, in the next step. This only requires to wait until all the columns have been processed before starting the propagation along the rows.

## Results

### Description of the algorithm

Based on the intuition that a path includes only vertical and horizontal edges, the algorithm computes the shortest path between one source and all the other nodes on an image with fewer instructions. Let us define the following variables

- H = height of the lattice

- W = width of the lattice

- si, sj = coordinates of the start node in the lattice (1 $le$ si ≤ H, 1 ≤ sj ≤ W)

- S = index of the start node = (sj-1)*H + si

- BED[i,j] : best distance to reach the node at coordinates (i,j), initialized to +inf

- PRED[i,j] : predecessor of the node at coordinates (i,j), initialized to null

- UPD[i,j] : matrix to keep node of which nodes have been updated. initialized to false.

The algorithm starts by updating the best distance of the starting node, BED[S] = 0 and the predecessor of S, PRED[S] = S. Now the iterative approach starts, doing at each iteration, for each column

- Select the nodes U[] whose best distance has been updated in the previous iteration.

- For each nodes u in U[]

- Compute the cost to reach any node x in the same column of u as
  c[x] = BED[u] +$w_{uv}$ +...+$w_{wx}$, which is the sum of the adjacent edges from u up to x (x above u, in the column), or the sum of the edges from u down to x (x below u, in the column).

- If c[x] ¡ BED[x] BED[x] = c[x]
    PRED[x] = x + 1 (x above u) or
    PRED[x] = x - 1 (x below u)
    UPD[x] = true
- Transpose the BED, PRED, and UPD matrices.

The algorithm stops when any update is done. Computational complexity In each column, the algorithm resembles the original Dijkstra algorithm, checking whether a node can be reached with less cost. For simplicity, let us consider a square lattice of size NxN. A column has $N$ nodes. Executing the naive implementation of the Dijkstra SSSP algorithm on a single column would lead to a cost of $O(\sqrt{N^2}) = O(N)$. This is repeated for all the columns, $K$ iterations (Figure 1). Therefore, the total computational complexity of the method is $O(k*n*\sqrt{n})$.

(Figure 1) reports the comparison of the proposed method with the classical implementation Dijkstra algorithm, showing the different ways in which costs are updated.

### Shortest path computation on images

Similarly to a regular grid, we consider a 2D image to be a graph with limited and fixed connectivity. More precisely, each pixel is connected to the four pixels which are closer in space (up, down, right, left).

By contrast to a regular grid, the cost of each edge depends on the brightness difference between two pixels.

Let us define an image as a matrix $I \in \mathbb{R}^{NxN}$. The cost of the edge connecting each pixel to its superior/inferior and left/right neighbor can be efficiently computed by optimized instructions for finite differencing, such as *diff* in Matlab. This yields to the creation of two matrices including the gradient components along orthogonal axes, thus the edge costs. In (Figure 2, C-D) we report the algorithm executed on two example images containing visual structures such as a text, and a flat logo. The algorithm exhibited a faster convergence as the shortest path in large areas with uniform color can be identified with two iterations.

### Anytime properties and convergence

The number of iterations $K$ vary according to the starting node and the graph. However, after a sufficient number of iterations, the algorithm finds shortest paths with the same cost of the original version, which coincide if unique shortest paths exist (Figure 2, A). Empirically, we found that the number of iterations depend on the number of turns that a path includes. $K = 1$ when a shortest-path in the solution is straight (i.e. does not include edges from different columns or rows). When an edge crossing two adjacent columns (or rows) is included, $K$ increases at least by 1. Empirically, on graphs with random edge cost (Figure 2, A-B, E), $K$ was proportional, but less, than $\sqrt{N}$. On graphs derived from images containing structures and areas with uniform intensity, the algorithm converged with less iterations (Figure 2, C-D). Therefore, the time required for running the algorithm is variable. However, parallel turning points are computed in parallel. Moreover, $K$ can be set to a maximum number of iterations. Indeed, the algorithm retains anytime properties. After the first two iterations (required to initialize the path-cost to all the nodes), at each iteration, the algorithm always computes a better solution and can be stopped at any time to obtain a sub-optimal solution. Convergence to the optimal solution is reached as demonstrated in Proof 1.

### Parallelism and cache friendly memory access

At each iteration, the proposed algorithm processes columns independently. When an iteration is completed, the algorithm reads the transposed matrix of edge costs. This can be computed before starting the algorithm, without the need for transposing at each iteration.

During an iteration, there is no need for communication between processes since each column reads and updates the values of the nodes (and edges) in the same column only. Therefore, parallelism is maximized inside each iteration. By contrast, each iteration has to wait until all the columns are processed. We benchmark the algorithm with respect to the classical implementation of the Dijkstra SSSP algorithm, showing a significant speedup both when processing columns sequentially (Figure 3, A - blue line) or gaining an additional 3.5x speedup for sufficiently large graphs by assigning different columns to different cores (Figure 3, A-B, green line).

Regarding the optimized memory access, let us assume that a matrix with a sufficiently high number of elements is organized on the main memory column-wise (Matlab/FORTRAN notation) on a system that uses a set-associative cache. When an element at position $i, j$ is accessed, the elements at position $i…i$+CL are loaded into the cache where CL is the size of cache line. Therefore, accessing those data will be much faster than accessing a random element of the matrix not in the cache.

As a consequence, successively accessing elements in columns on the matrix results in increased cache-hit than accessing elements row-wise or randomly.

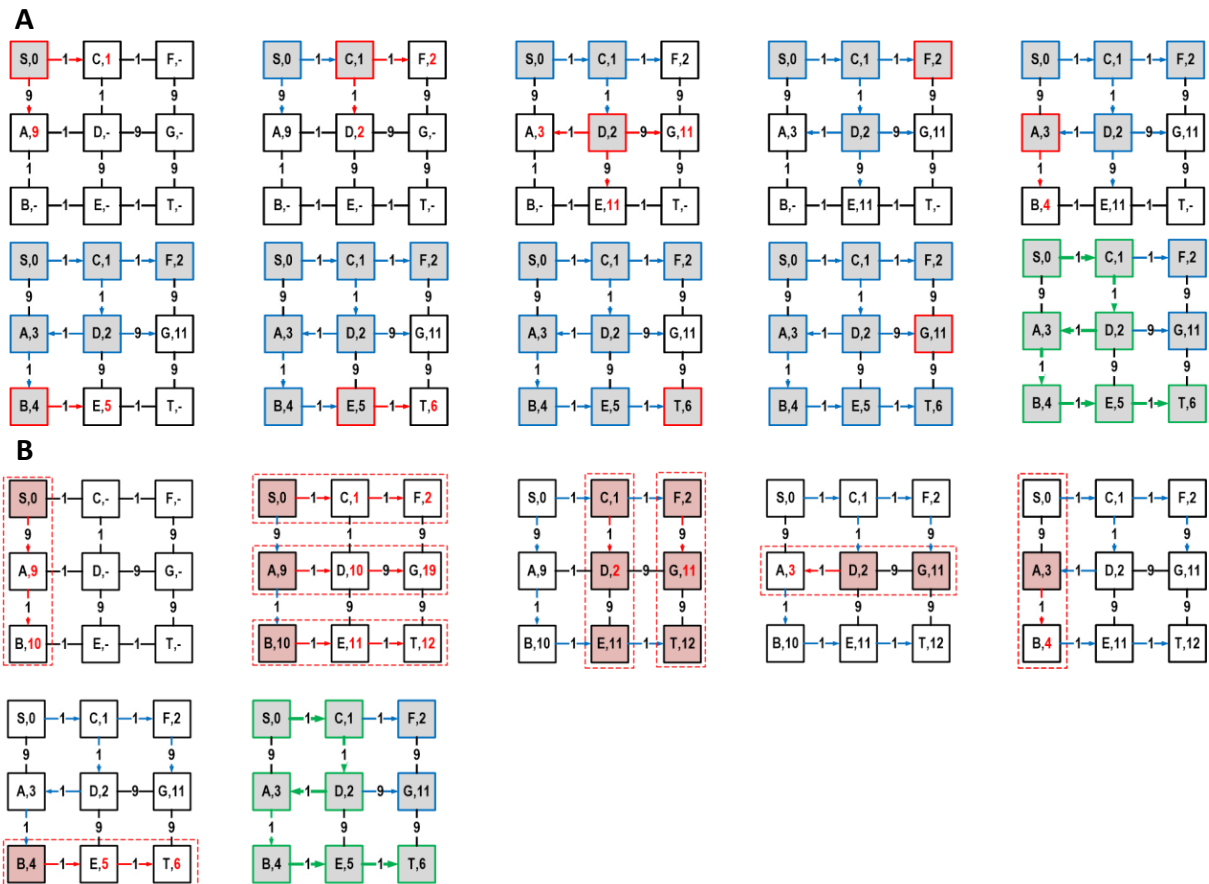

**Figure 1. Iterative implementation of the Dijkstra algorithm.** A. The classical algorithm selects the node reached with minimum cost and not yet selected. Then it updates the distance of the nodes connected to this node. B. Each column is processed independently, then the process is repeated on the rows and then on the columns.

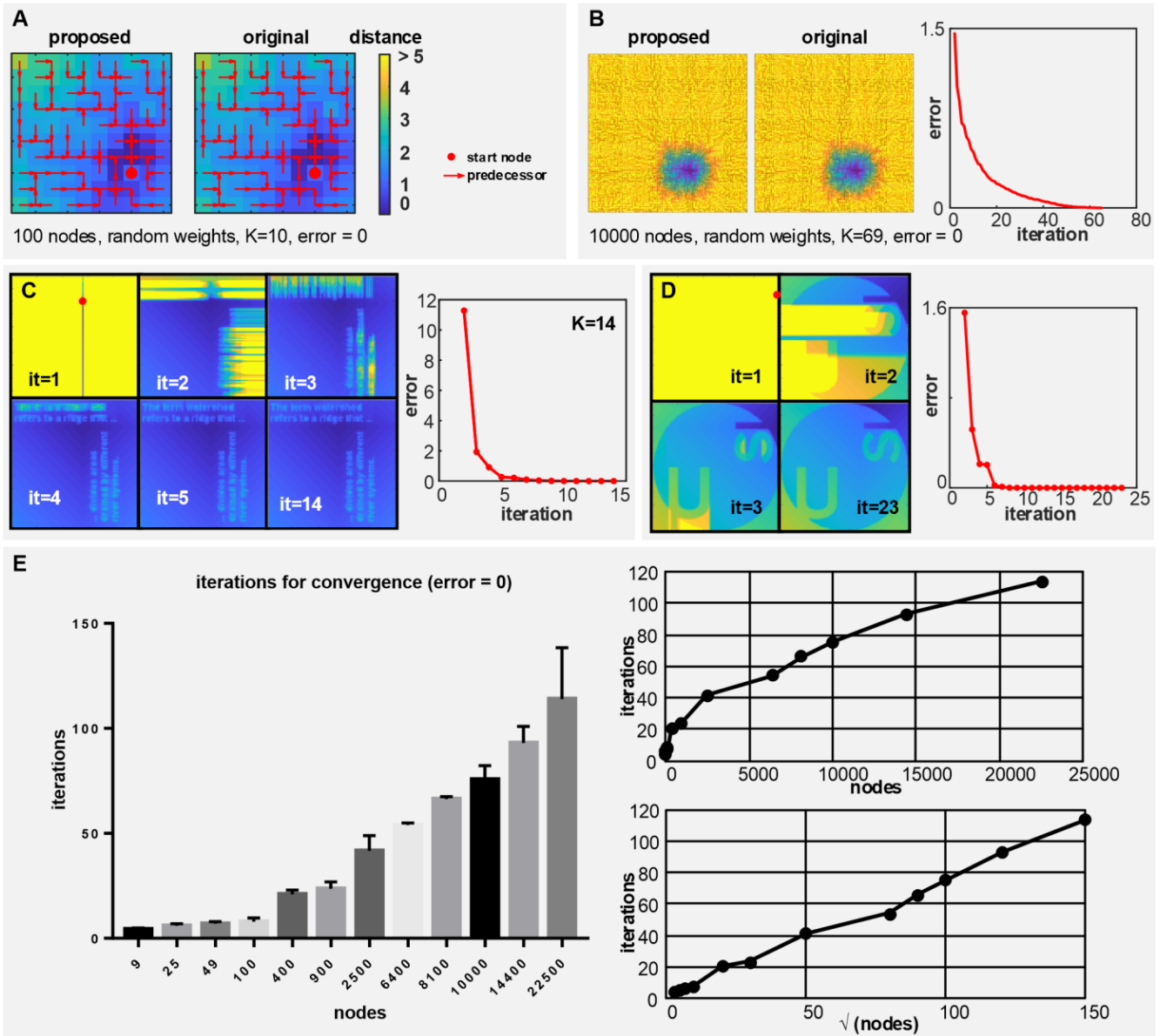

**Figure 2. Results of the iterative algorithm on different examples.** A. Color coded lattice of 100 nodes. Colors indicate the shortest-path distance from a random starting node (red dot). Red arrows indicate the predecessor of each node. Weights on the edges of the lattice are random. B. Color coded lattice with 10000 nodes and random edge weight, showing qualitatively no-difference between the proposed and the original algorithm (left), and a decreasing error at each iteration, which reaches 0 after 69 iterations (right). C. Results on an image containing structures (text) showing successive refinements of the shortest-path distance (left, color coded) and decreasing error at each iteration (right). D. Results on an image containing structures (USI logo) showing successive refinements of the shortest-path distance (left, color coded) and decreasing error at each iteration (right). E. Benchmark on lattice with different size and random edge weights, showing a number of iterations increasing with the size of the lattice, following a linear trend with the width (or height) of the lattice. (n = 3 replicas for each lattice size. Mean and SD are shown).

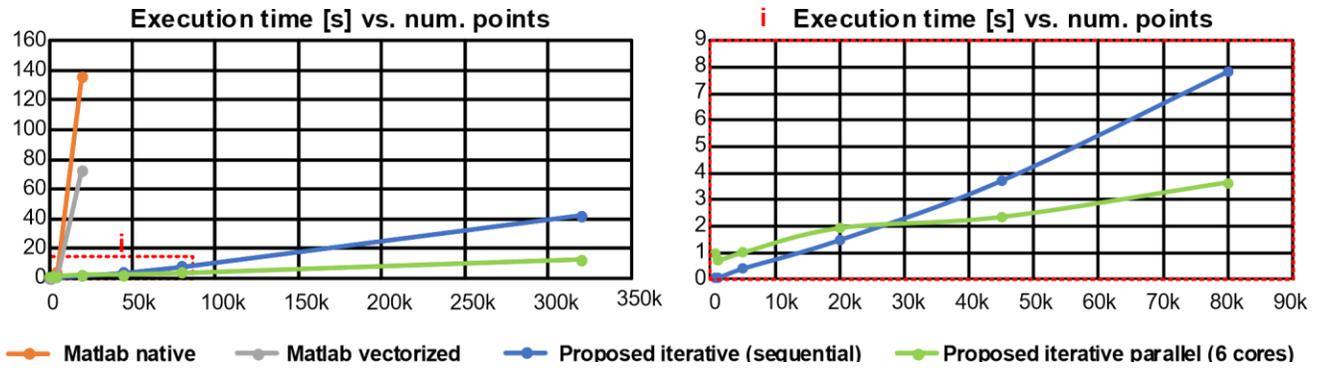

**Figure 3.** Benchmarking the iterative and parallel implementation of the Dijkstra algorithm A. Comparison of the proposed iterative method (blue: sequential, yellow: parallel) w.r.t. the classical implementation of the Dijkstra algorithm provided by the Matlab Bioinformatics toolbox (r2017b) and an omptimized implementation [7] using SIMD instructions. The proposed implementation allowed to tract image sizes that were poorly tractable by the classical implementation. B. Comparison of the proposed iterative methods on a machine with 6 cores (Intel Xeon E5-1650 v3), processing columns in sequential order (blue line) or in parallel by 6 threads (orange line). The parallel method achieves a 3.5x speedup when a sufficiently large graph is used.

### Proof of correctness

In this proof, we demonstrate that the path computed by the algorithm is the shortest path.

Let us define a lattice L4 ∈ G(V,E,w) as a graph with limited connectivity, in which an arbitrary node a ∈ V is connected to each of its upper, lower, left and right neighbors $b$ with an edge $(a,b) ∈ E$ having a non-negative cost $w_{a,b} ∈ R+$. An arbitrary path $\Gamma(s,t) ∈ L4$ is an ordered sequence of nodes connecting $s$ to $t$ $\Gamma = \{s,a,b,...,t\}$.

The cost of a path can be evaluated by means of a path-cost function $\xi : G \to R+$, $\xi(\Gamma) = c$.

For this proof, let us consider the following path-cost functions $\xi(\Gamma) = \sum_{(i,j)} w_{i,j}$ *with the edge (i,j)* ∈ $\Gamma$. However, the following demonstrations are valid for every non-decreasing path-cost function, as described in the previous section.

*Lemma 1. Every sub-path of a shortest path is a shortest path.*

Let us define a shortest path as $\Gamma^* = \{s,...,a,...,b,...t\}$ and assume ab absurdum that there exists another sub-path between the intermediate nodes $I(a,b)$, which is more convenient. If such a path existed, $\Gamma^*(s,t)$ would not be shortest. More formally, let $\Gamma^*(s,t)$ be the shortest path from a node $s$ to a node $t$ and let $\Gamma(s,t)$ be an arbitrary path between the same nodes. Since $\Gamma^*$ is optimal, $\xi(\Gamma^*) \le \xi(\Gamma)$ ∀$\Gamma$ ≠$\Gamma^*$. Let $I$ be a sub-path of $\Gamma^* = \{s,a,I*,b,t\}$ and let us assume that exists a sub-path $I ≠ I$ such that $\xi(I) < \xi(I^*)$.

Considering that the cost of an arbitrary path $\Gamma$ can be decomposed as $\xi(\Gamma) = w_{sa} + \xi(I) + w_{bt} = \xi(I) + K$     where $K >= 0$, derives that $\xi(\Gamma^*) = \min\xi(\Gamma) = \min(\xi(I) + K) = \min(\xi(I)) = \xi(I^*)$. Hence, $I$ is a shortest path.

*Corollary 1. Sub-paths of a shortest path including only vertical and horizontal edges are shortest-path.*

Let us consider the sub-paths of $\Gamma^*(s,t)$ that include only vertical or only horizontal edges. Each of these paths is a sub-path of a shortest path, therefore it is a shortest path. Hence, every shortest path on $L4$ can be decomposed into vertical and horizontal sub-paths, which are shortest-paths.

*Lemma 2. At iteration K, the algorithm finds the shortest paths containing at maximum K-1 turns.*

Let us consider an iteration as composed by both the propagation of path-costs in the vertical and in the horizontal directions. We refer to the vertical and horizontal steps of the k-th iteration as "kv" or "kh".

Let us also consider a starting node A and a target node T. The algorithm starts by assigning a path-cost to A = 0. This means that if the target node T = A, the algorithm already identified the solution at the first step. Since edge costs are non-negative, and path-cost functions non-decreasing, an other path would have a greater cost. Hence, all the shortest-paths with 0 edges will be identified at this step. By contrast, any other target node T ≠ A, would require additional iterations as the algorithm does not know yet the path nor the cost to reach the remaining nodes (Figure 4, dashed lines).

At the end of iteration 1v, the algorithm computes the path cost of the nodes in the same column of A, by summing the edge weights from A. This means that if a target node T is connected to A by means of a shortest-path including only vertical edges (T in the same column of A), such a path is found at iteration 1v. Similarly, at iteration 1h, the algorithms finds shortest paths from A to a target node T on the same row of A. This means that all the straight shortest-paths (with ¡ 1 turns) will be identified at the completion of iteration 1. Additionally, at iteration 1h the algorithm identifies L-shaped paths, which are formed by a vertical sub-path (in the same column of A) and a horizontal sub-path (in the same row of the target node). If these paths are optimal, they will not change at later iterations as their associated cost will be minimum. At iteration 2v, the algorithm compares if an arbitrary node, is reached with a lower cost following a vertical edge from a neighbor. Hence, it compares if the solution found at iteration 1h can be improved by substituting a horizontal edge, with a vertical edge, potentially introducing a turn. This means that the algorithm finds at the end of iteration 2, all the paths with ¡ 2 turns.

By induction, at the end of each iteration, the algorithm potentially substitute a vertical edge with a horizontal edge, introducing a turn if this yields a better solution. Hence it finds shortest paths with K-1 turns at completion of iteration K. This process is represented in Figure 4

*Proof. The algorithm finds the optimal combination of shortest sub-paths.*

Although a shortest-path can be decomposed in a sequence of shortest sub-paths including only vertical and horizontal edges (from now on, vertical and horizontal components), it remains to be demonstrated that the algorithm finds the optimal combination of these components.

Let us start demonstrating the correctness of the algorithm on a simplified lattice with 4 nodes where the shortest path from A to D has to be found (Figure ). Let us assume that the path A,B,D is the optimal one. At the first iteration, the algorithm finds that B is reached with a better distance, from A. At the second iteration, the algorithm finds that C is reached with a better distance from A, and D is reached with a better distance from B. At the third iteration, the algorithm finds that D is already reached with the best distance, as it could be reached from C with a worse distance. Focusing on the comparisons that the algorithm does, to reach the point D the algorithm compares the paths ACD and ABD, excluding ACD, which is the only other possible alternative.

On a lattice with an arbitrary number of nodes, a point P can be reached only by 4 neighbors (L,M,N,O). Considering that these four neighbors can be connected to the starting node with four paths (w,x,y,z), the algorithm compares at each iteration, the costs of the paths wLP, xMP, yNP, zOP. Hence, if the cost of a neighbor was improved in the previous iteration, the predecessor of P will be updated. For the Lemma 2, the cost of the predecessor is updated at iteration K-1, if and only if, a shortest-path with K-2 turns connects the predecessor to the starting node. Hence, after a sufficient number of iterations, the algorithm finds the shortest path and converges when the shortest path with the maximum numbers of turns have been identified.

## Discussion

The proposed algorithm allows computing the shortest path$\sqrt{}$ on images in $O(k*n*\sqrt{n})$ maximizing the sequential access to data and bringing a significant speedup with respect to other implementations.

As proposed in queue-based implementations of the Dijkstra algorithm, further optimization can be obtained by keeping a queue of the updated nodes at each iteration. This would allow exploring nodes only in the restricted neighborhood of the updated nodes in the next iteration.

The proposed method can generalize to N-dimensional images. A path in an N-dimensional represented as a hyperlattice, will be composed of edges parallel to one direction only. Hence, by selecting orthogonal directions at each iteration these paths can potentially be found. However, when dealing with anisotropic images, appropriate path-cost functions must be used. Indeed, as proposed in the previous section, a meaningful, non-decreasing path cost can be adopted for applying the proposed method to specific tasks.

## Methods

**Code and implementation** The proposed method was entirely written in Matlab (Mathworks), exploiting vectorial programming and associated SIMD instructions. It has been tested on an Intel Xeon E5-1650 v3 processor (6 cores, 3.50GHz, 384KB L1 cache, 1.5MB L2 cache, 15MB L3 cache) under Windows 8.1 operating system. Code is available on github at https://github.com/pizzagalli-du/anydijkstra under the MIT license.

## Acknowledgments

We are thankful to D. Morone, B. Thelen, G. Rovi, L. Karagyaur, A. Arini for technical discussion.

## Fundings

This work was supported by the Center for Computational Medicine in Cardiology (CCMC).
The authors declare no competing financial interests.

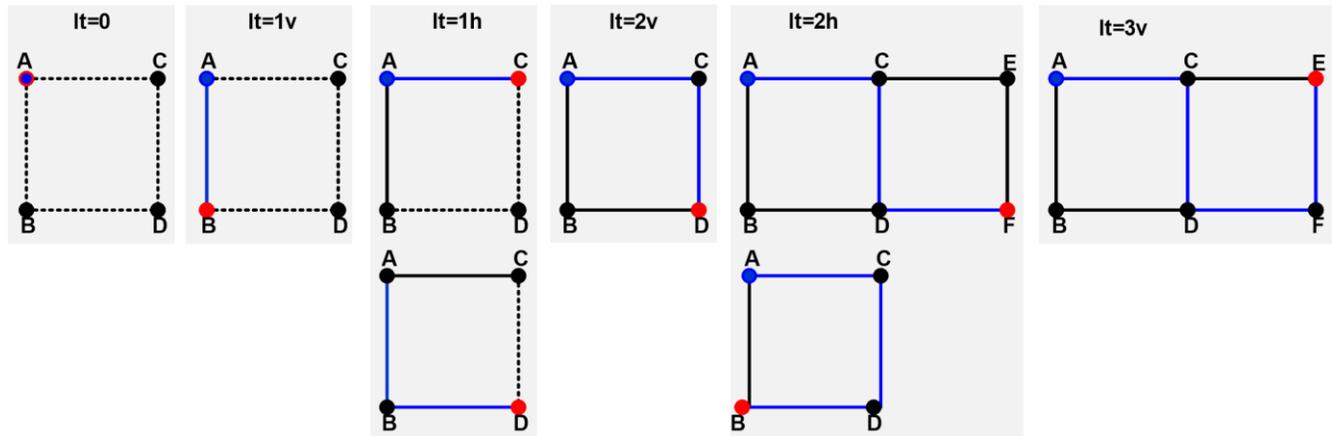

**Figure 4.** Types of shortest-paths found at each iteration At iteration *n*, the algorithms finds shortest paths with at maximum *n*−1 turns. It=nv or It=nh refer to the step in which distances are propagated vertically (v) or horizontally (h) at the n-th iteration. Blue node indicate the starting node. Red node indicate the target node. Dashed lines indicate edges which have not been evaluated yet at the current iteration. Continuous line indicate edges which have been already evaluated at the current iteration and are either not considered in the currently found shortest path (black) or considered (blue).